\documentstyle[preprint,aps]{revtex}
\input psfig
\begin{document}
\preprint{\small{Applied Physics Report 97-34. 
Submitted to Physica Scripta 980521}}
\title{Viscoelastic Acoustic Response of Layered Polymer Films
at Fluid-Solid Interfaces: Continuum Mechanics Approach}
\author{M. V. Voinova,$^{1,2}$\footnote{voinova@fy.chalmers.se}
 M. Rodahl,$^2$ M. Jonson,$^2$ 
and B. Kasemo$^2$}
\address{$^1$Department of Theoretical Physics, Kharkov State University,
 Kharkov 310077, Ukraine\\
$^2$Department of Applied Physics, Chalmers University of Technology and 
G\"oteborg University,\\
SE-412 96, G\"oteborg, Sweden}
\maketitle
\begin{abstract}

We have derived the general solution of a wave equation describing
the dynamics of two-layer viscoelastic polymer materials of arbitrary
thickness deposited on solid (quartz) surfaces in a fluid environment.
Within the Voight model of viscoelastic element, we calculate the acoustic response of the system to
an applied shear stress, i.e. we find the shift of the quartz
generator resonance frequency and of the dissipation factor, and show
that it strongly depends on the viscous loading of the adsorbed layers
and on the shear storage and loss moduli of the overlayers. These 
results  can readily be applied to quartz crystal acoustical 
measurements of the viscoelasticity of polymers which conserve their
shape under the shear deformations and do not flow ,
 and layered structures such as protein films adsorbed from solution 
onto the surface of self-assembled monolayers.



\end{abstract}
\pagebreak
\baselineskip=24pt
\section*{1. Introduction}

The analysis of viscoelastic behavior of biomolecular films at fluid-fluid
and fluid-solid interfaces is a very active interdisciplinary field
and a ``hot'' topic in the physics of modern surface phenomena, which has 
many possible applications [1-13]. Particular attention is focused now on 
macromolecular multilayer assemblies such as polymer ``brush" films [2], 
adsorbed protein layers [3,6,7-10] 
as well as on  self-assembled monolayers (SAM) [1,6,9]. The importance of 
studying such biomolecular sandwich structures arises from 
the prospect of their use as molecular biosensors and for medical and
enviromental applications [3,6,8-10]. The mechanical properties of these
structures can be determined with high accuracy from the frequency response
of a quartz oscillator in contact with them [6,8-14]. An increasing number
of theoretical and experimental investigations of quartz crystal microbalances
(QCM:s) functioning in liquid enviroments have opened a possibility for 
{\em in situ} measurements of biomolecular assemblies and for a quantitive
interpretation of their viscoelastic response [7-17]. Among theoretical
methods, a continuum mechanics approach is often used. In contrast to the 
electrical circuit method or a transmission line analysis
(\cite{41},\cite{42}, for a review see also [8] and Refs.~[13,15] therein), 
this approach is directly linked to the physical description of the system 
and provides a clear connection between measured characteristics and 
material parameters. 
Within
continuum mechanics the mechanical properties of viscoelastic materials are
usually related to energy storage and dissipation processes resulting 
from the balance between an applied stress and the ensuing relaxation 
processes in the material. The experimental conditions determine the amount
of stress, whereas the relaxation rates are intrinsic 
properties of the material [2,7,18,19].

The simplest way of accounting for the mechanical properties of a visco-elastic
material is to introduce a shear viscosity coefficient $\eta$ 
and a shear elasticity modulus $\mu$ within one of two basic
models due to (i) Maxwell and (ii) Voight. In contrast to the Maxwell model, 
the characteristics of the so called Voight element used in that model
is the following: It does not describe a 
flow at a steady rate; the viscoelastic element is described by complex shear modulus, real part
 of which (the storage modulus) is independent of frequency 
whereas its imaginary one (the loss modulus)
increases linearly with frequency [19].
In addition to the Maxwell and Voight models various combinations 
of them have been used (for a review, see Refs [14,18,2,7]).

The Maxwell model is usually applied to polymer solutions which can --- at
least for low shear rates --- demonstrate purely liquid-like behavior. The
Voight model is applicable for polymers which conserve their shape and do not
flow [18,19].

In the present study we analysed, within the continuum mechanics approach,
the viscoelastic response of two polymer layers covering a quartz plate
subject to shear deformations. The viscoelastic material was modelled as a
Voight element, with its parallel arrangement of a spring
and a dashpot. 
Even this relatively simple theoretical model has allowed us to demonstrate
 the
most important consequencies of adsorbate viscoelasticity in terms of a
deviation from the Sauerbrey relation \cite{41} relevant for a gas phase
environment and ultrathin layers. We have also
been able to elucidate the role of viscous loading when the oscillator 
is immersed in a bulk liquid.

\section*{2. Model}

We consider the case of two viscoelastic layers covering the surface of
a piezo-electric plate oscillating in a pure shear mode in a bulk liquid. 
We derive here a general solution of the corresponding wave equation and
analyze it using  ``no-slip" boundary conditions. In our 
model the result of Reed {\em et al.} [16] emerges as a special case
(i.e., when both layers are equivalent, and the upper (Newtonian) liquid 
has a vanishing density).

The geometry of the model system is shown in Fig.~1. We  treat the quartz
plate (index `0') as a harmonic oscillator whose resonant frequency in
vacuum [ 4, 5, 10-17], $f_0$, is
$$
f_0 = \frac{\ell_0}{2h_0}\sqrt{\frac{\mu_0}{\rho_0}} \, .
$$
Hence $f_0$ is a function of the thickness $h_0$ of the quartz plate, 
its density  $\rho_0$, and its elastic shear modulus $\mu_0$.

The quartz slab is treated as a loss-less generator when it operates in 
vacuum (or a gaseous environment). It is known that mass loading changes
the resonant frequency of the oscillator. For sufficiently thin non 
rigid overlayers, the shift of the resonant frequency is proportional
to the added mass only [4],
$$ \Delta f \approx f_0\Delta m \, ,
$$
and the energy dissipation due to viscous losses is then negligibly small
[6], i.e. such a layer demonstrates a solid-like response in vacuum or
gaseous experimental conditions.

In the opposite case of total immersion of the quartz slab into a liquid (in
general into a viscoelastic medium) both the energy dissipation due to the
oscillatory motion excited in the interfacial solid-liquid region, and the
resonant frequency shift, are functions of the overlayer  viscosity
(elasticity) and density [13]. Our results show that even very thin
viscoelastic films can dissipate a significant amount of energy when 
the quartz
plate with overlayers oscillates in the liquid phase.
The energy dissipation in QCM experiments can be obtained by measuring 
the
dissipation factor $D$ which is inversely proportional to the decay time
constant [6]
$$
D = \frac{1}{\pi f \tau} \, ,
$$
where $f$  is the resonant frequency. For a purely viscous bulk liquid, 
it has been shown
that the resonance frequency shift [15] and the dissipation factor
shift [5,6] depend on the density and shear
viscosity of the liquid.

Let us consider a viscoelastic overlayer medium within the Voight scheme of
using a viscoelastic element (Fig. 2) which consists of a spring and a 
dashpot in parallel. When a shear stress $\sigma_{xy}$ is applied 
to a Voight
element, the elastic response of the spring and the dashpot viscous 
resistance contribute to the stress/strain relation as follows:
\begin{equation}
\sigma_{xy} = \mu \frac{\partial u_x (y,t)}{\partial y} + 
\eta \frac{\partial v_x (y,t)}{\partial y} \, .
\end{equation}
Here  $u_x$  is the displacement in the x direction and the
corresponding velocity is $v_x$; $\mu$ is the elastic shear modulus 
and $\eta$ is the shear viscosity of the overlayer, respectively.

The wave equation for bulk shear waves propagating in a viscoelastic medium
is [16]:
\begin{equation}
\mu^{\ast}\frac{\partial^2 u_x(y,t)}{\partial y^2} = 
-\rho \omega^2 u_x (y,t) \, ,
\end{equation}
where $\mu^{\ast} \equiv \mu^{'}+i\mu{''} = \mu + i\omega \eta$ is a 
complex shear modulus.
The general solution of Eq.~(2) can be written as
\begin{equation}
 u_x(y,t) = \left( C_1 e^{-\xi y} + C_2 e^{\xi y}\right) e^{i\omega t} \, ,
\end{equation}
where
$\xi = \alpha + ik$
contains a decay constant $\alpha$,
\begin{equation}
\alpha = \frac{1}{\delta}\sqrt{\frac{\sqrt{1 + \chi^2} - \chi}{1 + \chi^2}}\\
\end{equation}
and a wave number $k$,
\begin{eqnarray}
k= \frac{1}{\delta}\sqrt{\frac{\sqrt{1 + \chi^2} + \chi}{1 + \chi^2}}
\nonumber \\
\chi = \frac{\mu}{\eta\omega}, \; \delta =\sqrt{\frac{2\eta}{\rho\omega}}.
\end{eqnarray}

Here we introduce the viscoelastic ratio  $\chi$ as the ratio between
real part (storage modulus) and imaginary one (loss modulus)
 of complex shear modulus and a viscous penetration
depth $\delta$, respectively.
The coefficients $C_1$ and $C_2$ in Eq.~(3) can be determined from 
the appropriate boundary conditions.

\section*{3. ``No-slip" conditions}
The so-called ``no-slip" boundary condition at the solid-overlayer
interface ($y = 0$)
corresponds
to a continuous variation of the displacement $u_x$ and of the shear
stress $\sigma_{xy}$ across the interface. These quantities
are related as follows,
\begin{eqnarray}
v_x = \frac{\partial u_x}{\partial t}, \; 
\sigma_{xy}&=&\mu_1 \frac{\partial u_x}{\partial y} +
\eta_1 \frac{\partial v_x}{\partial y}, \;
\nonumber \\
\sigma_{xy}&=&\mu_0 \frac{\partial q_x}{\partial y} ,~  q_x = u_x.\
\end{eqnarray}
Here $q_x$ correspondes to the component of quartz substrate 
displacement vector, $\mu_0$ is the substrate elastic shear modulus.

For the two-layer system we have a continuous variation at the 
surfaces $y = h_j$ of the velocities $v_x^{(i)}$  and the shear 
stresses $\sigma_{ik}^{i}$. Hence at
$y = h_1$ we have
\begin{eqnarray}
\label{3}
n_k \sigma_{ik}^{(1)}&=& n_k \sigma_{ik}^{(2)},
\nonumber \\
v_x^{(1)}&=&v_x^{(2)}, \qquad v_z^{(1)} = v_z^{(2)} = 0,
\nonumber \\
\sigma_{ik}^{(1)}&=&-p_1 \delta_{ik} + \eta_1(\frac{\partial {v_i}^{(1)}}
{\partial x_k} + \frac{\partial {v_k}^{(1)}}{\partial x_i})
+ \mu_1 \left( \frac{\partial u_i^{(1)}}{\partial x_k} 
+ \frac{\partial u_k^{(1)}}{\partial x_i}\right),
\\
 \sigma_{ik}^{(2)}&=&-p_2 \delta_{ik} + \eta_2(\frac{\partial {v_i}^{(2)}}
{\partial x_k} + \frac{\partial {v_k}^{(2)}}{\partial x_i})
+ \mu_2 \left( \frac{\partial u_i^{(2)}}{\partial x_k} 
+ \frac{\partial u_k^{(2)}}{\partial x_i}\right),
\nonumber
\end{eqnarray}

The no-slip boundary condition at $y =h_2$ gives:
\begin{eqnarray}
n_k \sigma_{ik}^{(2)}&=& n_k \sigma_{ik}^{(3)},
\nonumber \\
v_x^{(2)}&=&v_x^{(3)}, \qquad v_z^{(2)} = v_z^{(3)} = 0,
\nonumber \\
\sigma_{ik}^{(2)}&=& -\rho_2 \delta_{ik} + \eta_2(\frac{\partial {v_i}^{(2)}}
{\partial x_k} + \frac{\partial {v_k}^{(2)}}{\partial x_i})
+ \mu_2 \left( \frac{\partial u_i^{(2)}}{\partial x_k}
+ \frac{\partial u_k^{(2)}}{\partial x_i}\right),
\\
\sigma_{ik}^{(3)}&=& -\rho_3 \delta_{ik} + \eta_3(\frac{\partial {v_i}^{(3)}}
{\partial x_k} + \frac{\partial {v_k}^{(3)}}{\partial x_i}) \, .
\nonumber
\end{eqnarray}

Finally, at the free surface, $y =h_3$, the boundary condition is
\begin{equation}
\eta_2 \frac{\partial v^{(3)}_x}{\partial y} = 0 \, .
\end{equation}
A general solution for the wave equation (2) with boundary conditions
(6-9) can be written as
\begin{equation}
v_x = v_0 \frac{e^{2\xi_1 y} + Ae^{2\xi_1 h_1}}{e^{\xi_1 y}\left( 
1 +  Ae^{2\xi_1 h_1}\right)},
\nonumber
\end{equation}
\begin{equation}
A = \frac{\kappa_1\xi_1 \left( 1+ le^{2\xi_2 \Delta h_1}\right) -
\kappa_2\xi_2 \left( 1 - le^{2\xi_2 \Delta h_1}\right)}
{\kappa_1\xi_1 \left( 1+ le^{2\xi_2 \Delta h_1}\right)+ 
\kappa_2\xi_2 \left( 1 - le^{2\xi_2 \Delta h_1}\right)}, \;
l = \frac{\kappa_2\xi_2 + \kappa_3\xi_3 \tanh{\left(\xi_3 \Delta h_2\right)}}
 {\kappa_2\xi_2 - \kappa_3\xi_3 \tanh{\left(\xi_3 \Delta h_2\right)}}
\nonumber
\end{equation}
\begin{equation}
\Delta h_1 = h_2 - h_1, \; \Delta h_2 = h_3 - h_2,
\end{equation}
\begin{equation}
\xi_{1,2} = \sqrt{-\frac{\rho_{1,2}\omega^2}{\mu^{\ast}_{1,2}}} =
\alpha_{1,2} + ik_{1,2}, \;
\xi_3 = \sqrt{i\frac{\rho_3 \omega}{\eta_3}}
\nonumber
\end{equation}
\begin{equation}
\kappa_{1,2} = 
\eta_{1,2} - \frac{i\mu_{1,2}}{\omega} \equiv \frac{\mu^{\ast}_{1,2}}{i\omega}, \; \:
\kappa_3 = \eta_3 \, .
\nonumber
\end{equation}

The  shift in resonance frequency
 $\Delta f$  and dissipation factor $\Delta D$  
are obtained, respectively, from the imaginary and 
real parts of the $\beta$-function [6]:
\begin{eqnarray}
\Delta f = {\rm Im}\left( \frac{\beta}{2\pi\rho_0 h_0} \right), 
\nonumber \\
\Delta D = -{\rm Re}\left( \frac{\beta}{\pi f \rho_0 h_0} \right)
\end{eqnarray}
where
\begin{equation}
\beta = \kappa_1 \xi_1 \frac{1-Ae^{2\xi_1 h_1}}{1+Ae^{2\xi_1 h_1}} \, .
\end{equation}

Below we analyze the general results expressed in Eqs.~(10-16) 
in the limit cases of thin and thick
viscoelastic layers deposited on the quartz oscillator surface when the
oscillator operates (i) in vacuum (or a gaseous
environment) and (ii) in a bulk liquid.

\section*{4. Results}
\subsection*{4.1. Viscoelastic layer in vacuum}
From general solution for the wave equation (10) it follows that the
overlayer thickness always appears as a dimensionless $h \xi$ combination.
Thus, the limit cases of "thin"/"thick" film correspond to the situations
wheather the film thickness is much smaller/greater in comparison with
the reverse value of a decay $\alpha$ constant and the propagation $k$ 
constant which is reduced to the following criteria:

in the 'thin film' approximation  $h/\delta \ll \sqrt{\chi/2}$ for 
$\chi > 1$
and  $h/\delta \ll 1-\chi/2$   for $\chi < 1$;

'thick layer' correspondes to inequalities $h/\delta \gg \chi\sqrt{2\chi}$
 for $\chi > 1$ and $h/\delta \gg 1+\chi/2$ for $\chi < 1$.

For pure viscous film this criterium corresponds to the well-known
 ratio between a film thickness $h$ and a viscous penetration 
depth $\delta$: in the thin layer approximation $h/\delta\ll 1$ [20].

Let us  consider the solution of wave equation for one viscoelastic layer
on the QCM surface when the system oscillates in vacuum (or in a gas).
In such case, from general expressions (11-16) we get for $\beta$-function: 
\begin{equation}
\beta = -\kappa \xi \tanh{\left( \xi h \right)} \, .
\end{equation}
By a series expansion on $h\alpha\ll 1$ and $h k \ll 1$ in special
case of thin overlayer, we find from (15) and (17) that
\begin{equation}
\Delta f  \approx -\frac{1}{2\pi\rho_0 h_0}h\rho\omega (1
 - \frac{2h^2 }{3\delta^2(1+\chi^2)})\\
\end{equation}
\begin{equation}
\Delta D \approx \frac{2h^3\rho\omega }{3\pi f\rho_0 h_0} \frac{ 
\chi}{\delta^2(1+\chi^2)},~ \chi = \frac{\mu}{\eta\omega}, ~
 \delta =\sqrt{\frac{2\eta}{\rho\omega}}\\
\end{equation}

 It is clear from Eqs.~(18) and (19) that to linear order
in the (small) thickness, the dissipation factor  vanishes, while the
 frequency shift is a
function of oscillation
frequency and mass of the layer. This is the well-known Sauerbrey
relation for the loss-less quartz oscillator overlayered with a thin film:
$$
\Delta f_{Sauerbrey} = - \frac{f_0}{\rho_0 h_0}\Delta m,  \;
~\Delta m \equiv \rho h
$$
$$
\Delta D =0
$$
The viscosity and elasticity of the overlayer only appears in the third order
approximation with respect to thickness. For sufficiently thin layers, it 
leads to a small correction to the Sauerbrey relation which can be hardly
observed experimentally.

In the opposite case of a "thick" viscoelastic layer $(h \alpha \gg 1, \,
hk \gg 1)$
we readily find that $\beta \approx -\kappa \xi$ and that
\begin{eqnarray}
\Delta f 
\approx -\frac{1}{2\pi\rho_0 h_0} \sqrt{\frac{\rho}{2}} \left\{ \eta\omega
\sqrt{\frac{\sqrt{\mu^2 + \eta^2\omega^2} + \mu}{\mu^2 + \eta^2\omega^2}}
-\mu \sqrt{\frac{\sqrt{\mu^2 + \eta^2\omega^2} - \mu}{\mu^2 + \eta^2\omega^2}}
\right\}
\\
\Delta D 
\approx \frac{1}{\pi f\rho_0 h_0} \sqrt{\frac{\rho}{2}} \left\{ \eta\omega
\sqrt{\frac{\sqrt{\mu^2 + \eta^2\omega^2} - \mu}{\mu^2 + \eta^2\omega^2}}
+\mu \sqrt{\frac{\sqrt{\mu^2 + \eta^2\omega^2} + \mu}{\mu^2 + \eta^2\omega^2}}
\right\} \, .
\end{eqnarray}
The shift of resonance frequency (20) and the dissipation factor changes (21)
are functions of film viscoelasticity and can be measured in QCM experiments.

In the viscous limit $(\mu \rightarrow 0)$ we recover the result of 
Ref.~[15]
for the frequency shift,
\begin{equation}
\Delta f \approx - \frac{1}{2\pi\rho_0 h_0}
\sqrt{\frac{\rho\eta\omega}{2}} \, ,
\end{equation}
and the result of Refs.~[5,6] for the shift of the dissipation factor,
\begin{equation}
\Delta D\approx \frac{1}{\rho_0 h_0}\sqrt{\frac{2\rho\eta}{\omega}} \, .
\end{equation}

\subsection*{4.2. Viscoelastic layers in a bulk liquid}

Let us consider  two thin viscoelastic overlayers of thickness
 $h_j \,(j = 1,2)$ under a bulk Newtonian liquid for which its thickness
 $\Delta h_2$ satisfies the inequality
$\xi_3 \Delta h_2 \gg 1$ . 
In this case we may expand the expression for $\beta$ 
and keep only terms linear in $h_j  \alpha_j \ll 1$ and 
 $h_j k_j \ll 1$. The acoustic response of the  QCM with the thin 
viscoelastic layers immersed in a Newtonian bulk liquid (index ``3"):       
\begin{eqnarray}
\Delta f \approx -\frac{1}{2\pi\rho_0 h_0}\left\{ \frac{\eta_3}{\delta_3} +
\sum_{j=1,2}\left[ h_j \rho_j\omega - 2h_j\left(\frac{\eta_3}{\delta_3}\right)
^2 \frac{\eta_j\omega^2}{\mu_j^2 + \omega^2\eta_j^2}\right]\right\}\\
\Delta D \approx \frac{1}{2\pi f \rho_0 h_0}\left\{ \frac{\eta_3}{\delta_3} +
\sum_{j=1,2}\left[ 
2h_j\left(\frac{\eta_3}{\delta_3}\right)
^2 \frac{\mu_j\omega}{\mu_j^2 + \omega^2\eta_j^2}\right]\right\} \, .
\end{eqnarray}
From Eqs.~(24) and (25) it follows that for ultrathin films the
contribution of the film is small in comparison with the bulk
liquid acoustic response 
$\Delta f \sim -\eta_3/\delta_3, \; \Delta D \sim \eta_3/\delta_3$.
However, a thin layer with a
finite thickness will demonstrate a different acoustic response
depending on the ratio between the viscosity and the elasticity 
of the film.

To make a picture more clear,
let start from only one thin viscous overlayer under a bulk Newtonian
liquid. For a purely viscous overlayer, the resonant frequency shift
 depends on a bulk $\eta_3/\delta_3$ term and the
 difference between the film mass contribution $(h_1\rho_1\omega)$
and its viscous contribution $(\sim h_1/\eta_1)$ multiplied by 
factor $(\eta_3/\delta_3)^2$  due to the penetration
of acoustical vibrations through the adsorbed layer into a bulk liquid
and corresponding energy loss due to the viscous dissipation.
The dissipation factor  (25) is determined by properties
of bulk liquid.

 We found, that even in this simple case there exists an
important for experiments situation when the resonance frequency
shift of layered system (24) is equal to a bulk liquid responce only
but not of adsorbed layer one. This occurs for such a relationship
 between viscosities and densities of the media, i.e.
 $\eta_3 \rho_3 = \eta_1 \rho_1$, at which 
the 'mass responce' will compensate the viscous one.
We can explain this subtraction as a result of damping and reflection of
 travelling acoustic waves in a layered medium with 
different density and viscosity of the layers.  

In the opposite case,  for purely elastic film $(\eta_1 \rightarrow 0)$ 
one can find that only the dissipation
factor change depends on the film elasticity $(\Delta D \sim h_1/\mu_1)$, 
while the frequency
shift contains the mass response of the film $(\Delta f \sim h_1\rho_1)$
but not the elastic one.
It is easy to see that for a gaseous environment when $\eta_3$ 
tends to zero, we recover the  Sauerbrey relation (see also the next
 paragraph). Thus, it is 
essential to note that the viscous bulk loading of the adsorbed layer
leads to the apparence of a viscous (viscoelastic) contribution
of the layer and hence opens up a possibility for the corresponding QCM
measurements to determine the viscoelastic parameters 
of the overlayer(s) when the resonator oscillates in a bulk liquid. 

In Figs. ~3 - 6 we have presented the calculated change in resonance frequency
and in the dissipation factor as a function of the viscoelastic film
thickness $h_1$ for a given value of the layer viscoelasticity 
($\mu_1= 10^4$ - $10^5$ $N/m^2$, 
$\eta_1=0.01$ $Ns/m^2$, $\rho_1=10^3$ $kg/m^3$) when a
quartz plate oscillates with frequency $f_0=5$ $10^6$ Hz in bulk water 
($\rho_3 = 10^{3}$ kg/m$^3$, $\eta_3= 10^{-3}$ $kg/m^3$). 

In real experimental situation, not only changes of the thickness of
 adsorbed layer must be taken into account. Variations  of
$\Delta f(\mu , \eta)$ and $\Delta D (\mu , \eta)$ during the adsorption
 can be of crucial importance  as viscoelastic 
parameters of the adsorbate can also vary during the process of coverage.

To illustrate the variation of $\Delta f$ and $\Delta D$  with the 
viscoelastic layer
parameters  $\mu$ and $\eta$, we have shown in Figs.~7 - 8
the results of a
numerical simulation of the model for the thin viscoelastic layer of 
a given thickness
 $h_1 = 10^{-6}$m and density $\rho_1 = 10^{3}$ kg/m$^3$, when a generator
oscillates with frequency $f_0$ = $10^7$ Hz in bulk water.
The resulting nonmonotonic behavior of
$\Delta f(\mu , \eta)$ and $\Delta D (\mu , \eta)$  gives the evidence
for key contribution of viscoelasticity  in complex 
polymer adsorbed layers dynamics in a liquid medium.

\section*{5. Conclusions}
We have found a general solution of the wave equation describing the 
propagation and
decay of shear waves in layered viscoelastic medium in contact with a
quartz surface. This sandwich structure can be used to 
model the response of viscoelastic polymer and  biomolecular films such as polymer ``brush"
films or proteins adsorbed from solution onto a solid substrate
modified by SAM covering before adsorption.
Our results can be summarized as follows.

(i) We have analysed both ''thin" and ''thick" layer acoustical responces.
These two limit cases correspond to wheather the film thickness is much 
smaller/greater in comparison with the viscous penetration depth
 $\delta=\sqrt{2\eta/\rho\omega}$
and the viscoelastic ratio $\chi = \mu/\eta\omega$.

(ii) To elucidate the validity of the Sauerbrey ``ideal mass thin layer"
 response,  we have calculated the ''thin"  viscoelastic layer resonance
 frequency shift.
We found that  the  ``ideal mass layer"
response corresponds to a purely elastic behavior of the Voight element,
i.e. to an element with zero shear viscosity. Due to the viscosity of the
adsorbed layer, the acoustical responce of the system becomes to diverge
with an ``ideal mass thin layer" behavior when the ratio of the film
 thickness $h$ and the penetration depth $\delta$ tends to $1/2$ (for small values of viscoelastic ratio $\chi\ll 1$).

(iii) In case of a ''thick" viscoelastic medium we found the shifts in
resonance frequency and dissipation factor as functions of the ratio 
of storage and loss moduli. Using Eqs.~(20) and (21) it is possible to 
calculate these moduli using data from the corresponding acoustic experiments.

(iv) We derived the general expression for the acoustic response of two
 viscoelastic
layers of arbitrary thickness covering the surface of a quartz oscillator
 when it is
immersed in a Newtonian liquid. This result allows us to analyze
the dynamics of a layered medium with arbitrary layer thicknesses
and with a viscoelastic ratio which can vary during the coverage.

 We have presented analytical expressions
for the resonance frequency shift and the dissipation factor response of 
 thin
viscoelastic layers in contact with a liquid. The analysis showed that in
contrast to oscillations in vacuum or in a gas environment, even a
thin viscoelastic layer will dissipate a significant amount of energy. 
We have therefore also shown that a viscous bulk loading
of the overlayer opens up the possibility to measure its 
viscoelastic properties from simultaneous measurements of 
$\Delta f$ and $\Delta D$ using a QCM operating in a liquid enviroment.

The results of the model can readily be applied
to quartz crystal acoustic measurements of adsorbed proteins
which conserve their shape during adsorption and do not flow under
shear deformation and of the polymer films far from the glass
 transition region.

\section*{ Acknowledgements}

One of the authors (MV) thanks  Dr. C.-Y.~D.~Lu 
(Cavendish Laboratory, Cambridge) and
Dr. Bo~Persson (J{\"u}lich) for fruitful discussions and
 for helpful correspondence.
This work was supported by the Royal Swedish Academy of Sciences 
(KVA) and the Swedish TFR.

\begin{figure}
\centerline{\psfig{figure=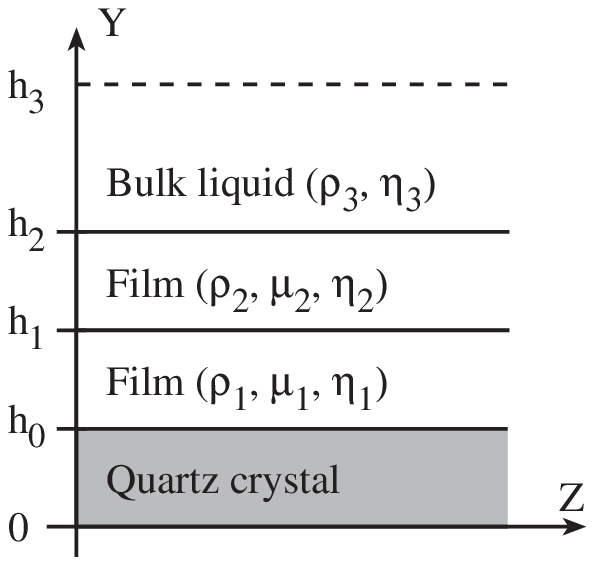,width=10cm}}
\vspace{1.0cm}
\caption{ Geometry of a quartz crystal microbalance (QCM) 
 covered by a double-layer viscoelastic film.
The QCM system oscillates in a bulk liquid.}
\label{fig1}
\end{figure}
\newpage

\begin{figure}[h]
\centerline{\psfig{figure=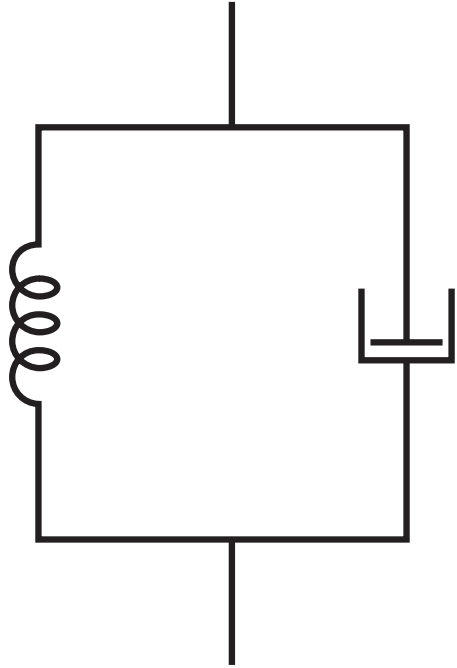,width=6cm}}
\vspace{1.0cm}
\caption{ Schematic depiction of the Voight viscoelastic
 element.} 

\label{fig2}
\end{figure}
\newpage

\begin{figure}[h]
\centerline{\psfig{figure=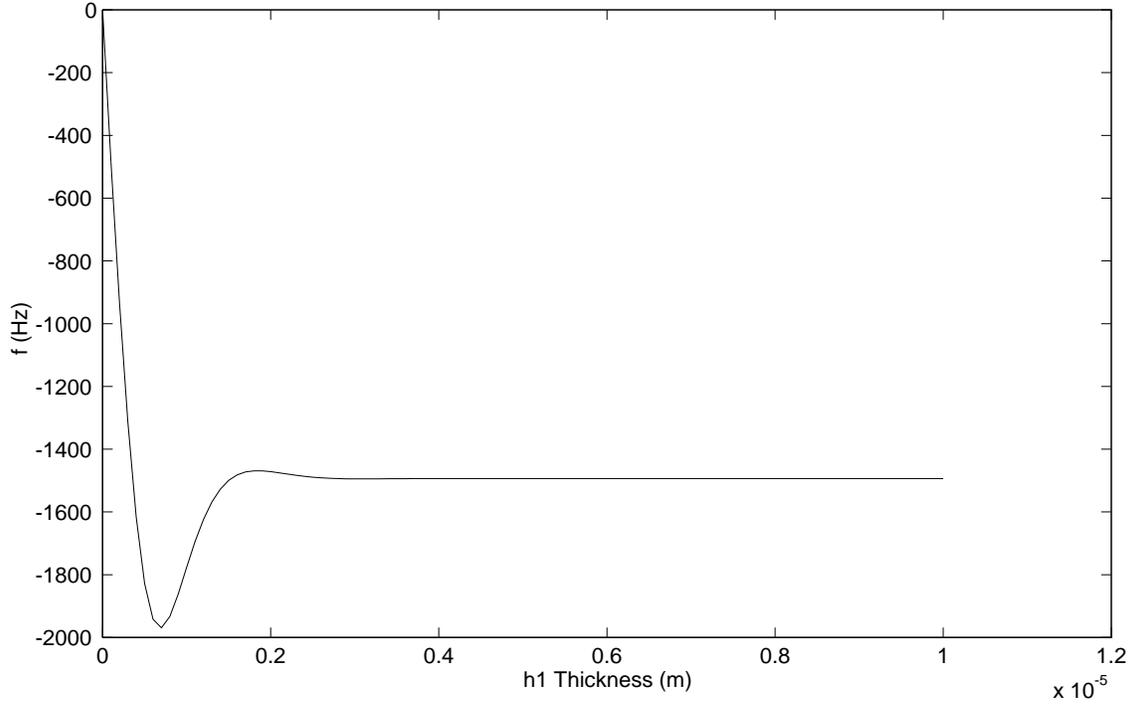,width=15cm}}
\vspace{1.0cm}
\caption{ Numerically calculated resonance frequency shift as a function 
of a viscoelastic overlayer thickness $h_1$; $\rho_1$ = 1000 kg/m$^3$,
$\eta_1$ = $10^{-2}$ Ns/m$^2$, $\mu_1$ = $10^4$ N/m$^2$.
 The QCM oscillates in bulk water; $\rho_1$ = $10^{3}$ kg/m$^3$,
$\eta_3$ = $10^{-3}$ Ns/m$^2$,
 $f_0$ = $5\cdot 10^6$ Hz.} 

\label{fig3}
\end{figure}
\newpage
\begin{figure}[h]
\centerline{\psfig{figure=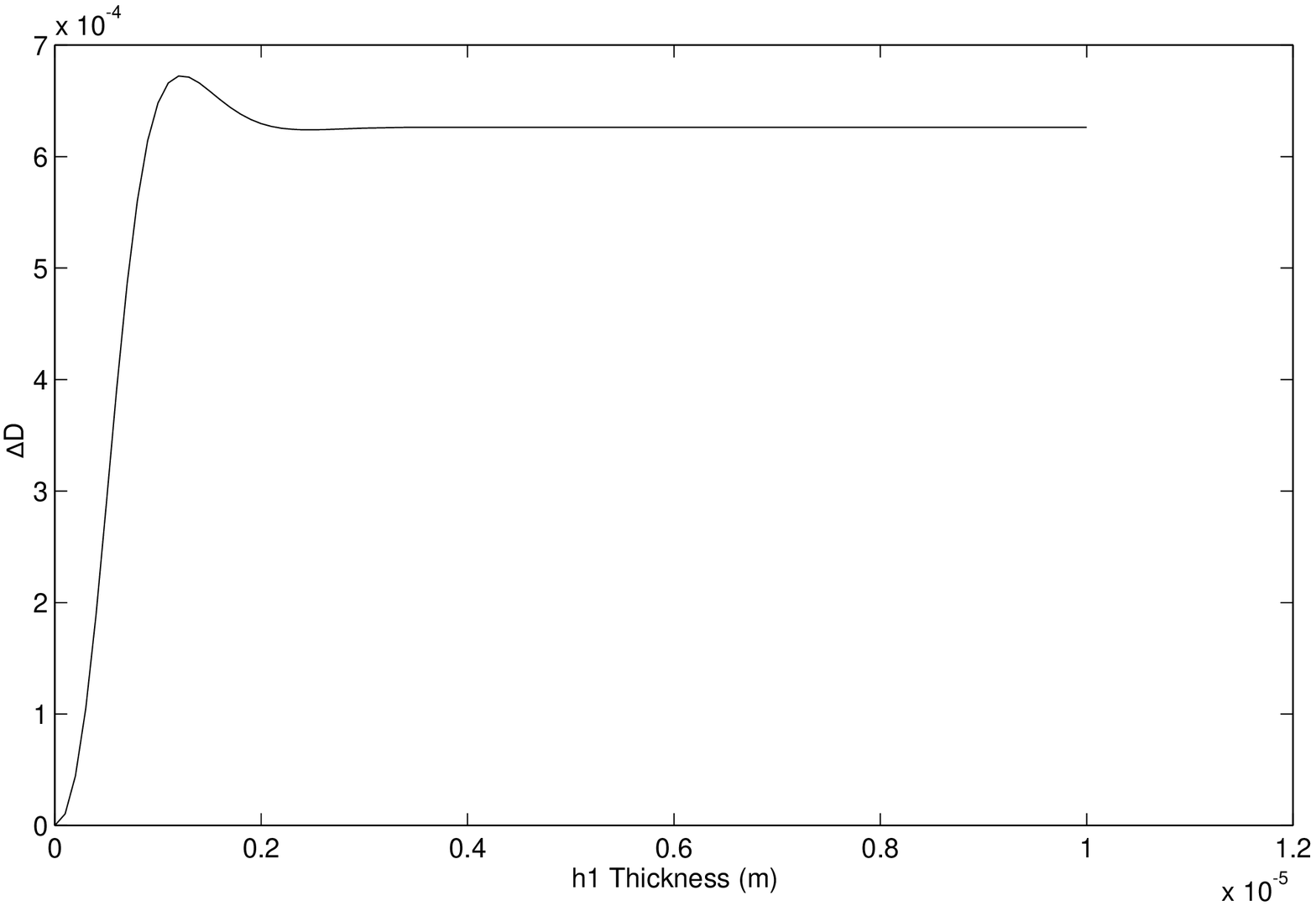,width=15cm}}
\vspace{1.0cm}
\caption{ Numerically calculated dissipation factor change as a function
of a viscoelastic overlayer thickness (for the same parameters as in Fig.~3).
 The QCM oscillates in bulk water.}
          
\label{fig4}
\end{figure}
\newpage
\begin{figure}[h]
\centerline{\psfig{figure=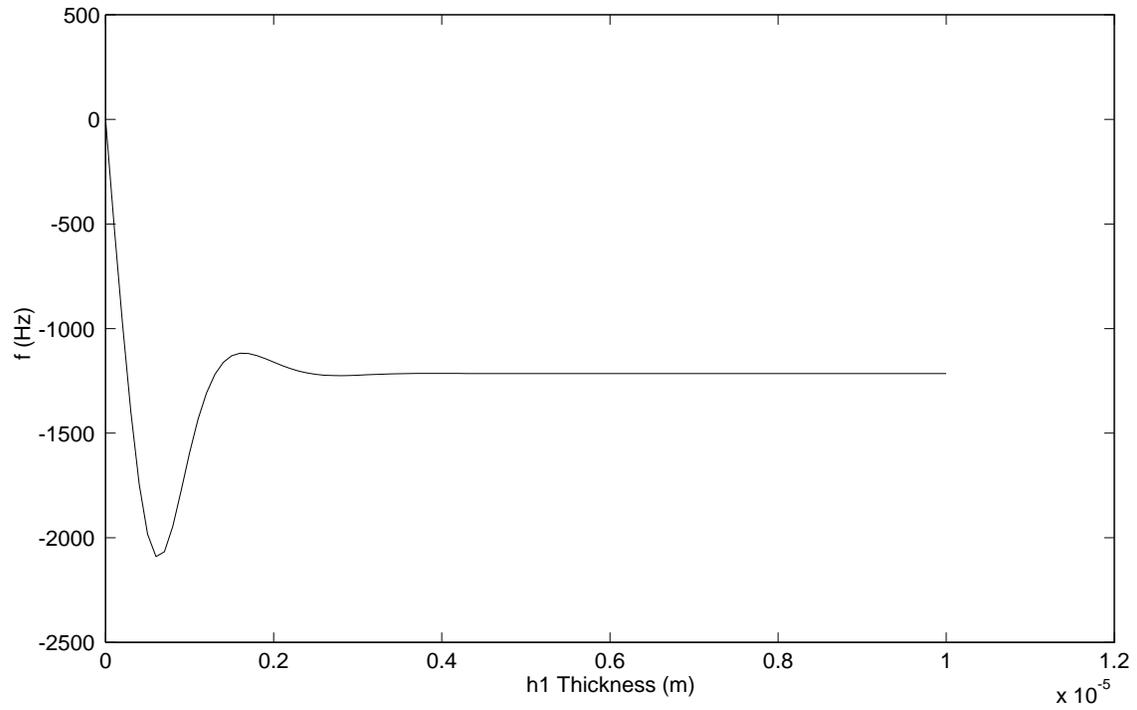,width=15cm}}
\vspace{1.0cm}
\caption{ Numerically calculated resonance frequency shift as a function 
of a viscoelastic overlayer thickness $h_1$; all the 
overlayer parameters are the same as in Fig.3 but with 
$\mu_1$ = $10^5$ N/m$^2$.
 The QCM oscillates in bulk water; $f_0$ = $5\cdot 10^6$ Hz.} 

\label{fig5}
\end{figure}
\newpage
\begin{figure}[h]
\centerline{\psfig{figure=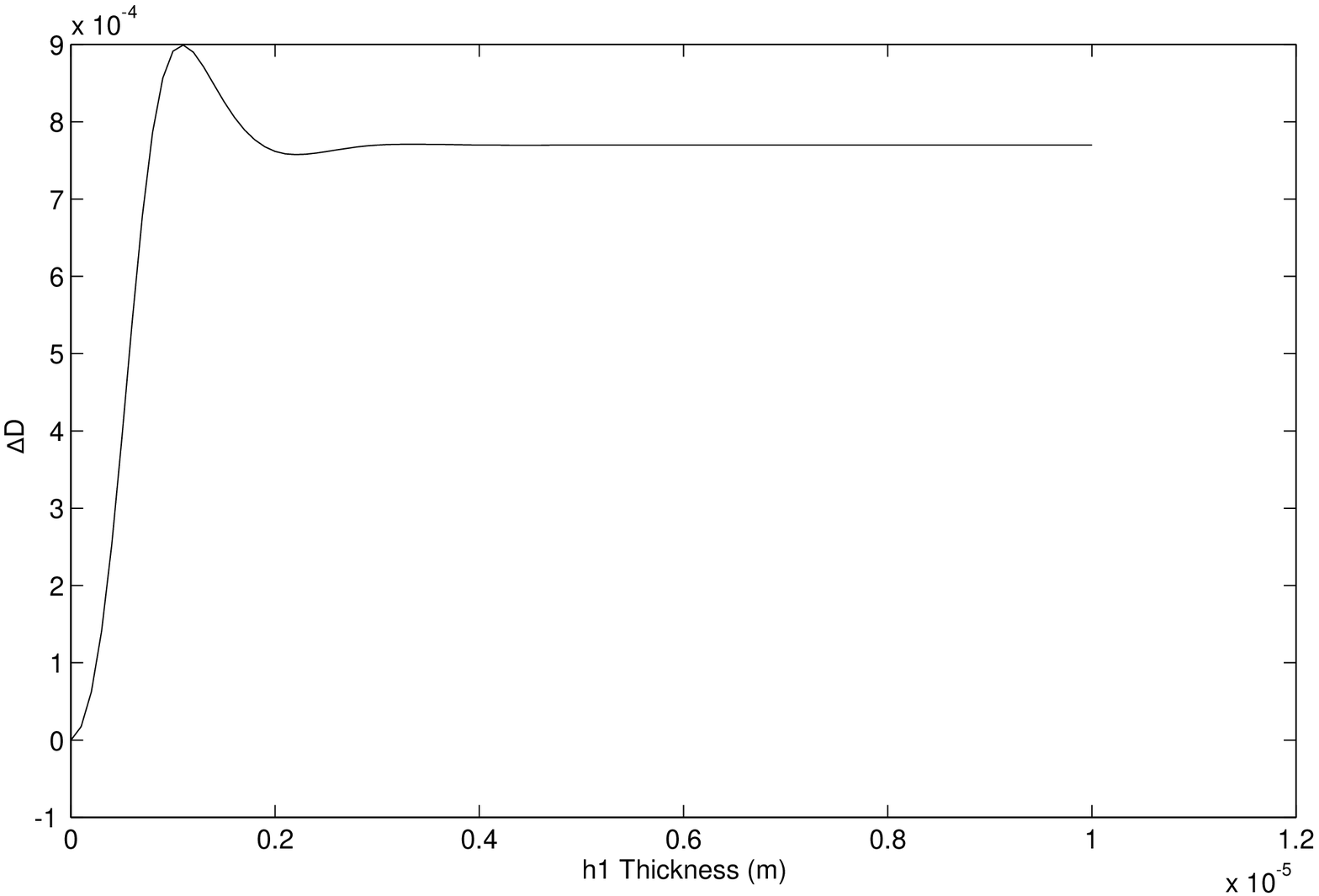,width=15cm}}
\vspace{1.0cm}
\caption{ Numerically calculated dissipation factor change as a function
of a viscoelastic overlayer thickness (for the same parameters as in Fig.~5).
 The QCM oscillates in bulk water.}
          
\label{fig6}
\end{figure}
\newpage
\begin{figure}[h]
\centerline{\psfig{figure=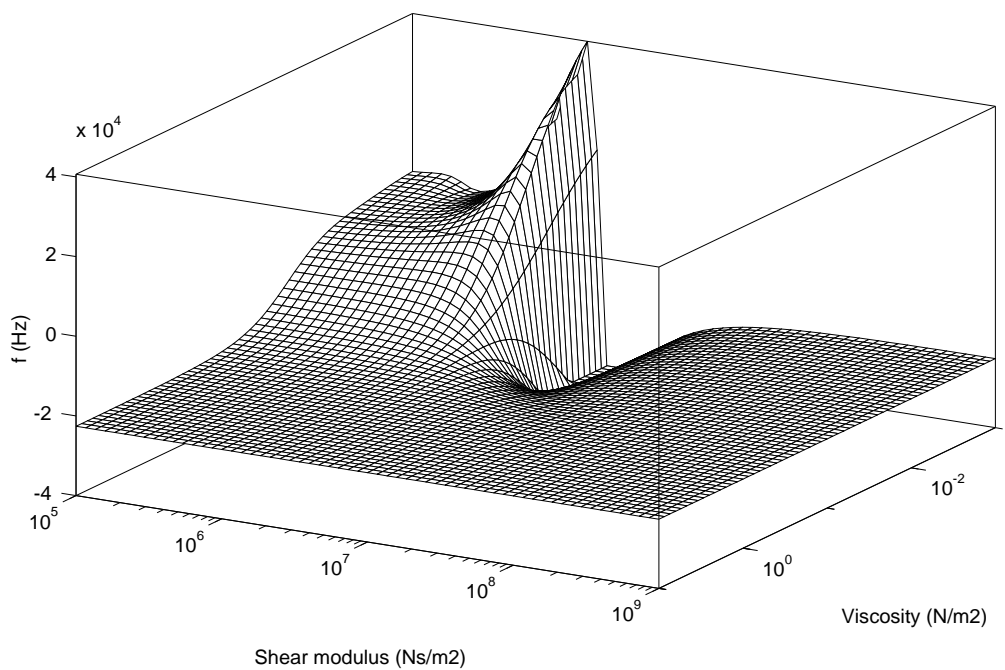,width=15cm}}
\vspace{1.0cm}
\caption{ Numerically calculated resonance frequency shifts as a function of
the shear elasticity modulus $\mu_1$ and the shear viscosity $\eta_1$ of 
the viscoelastic overlayer of thickness  $h_1 = 10^{-6}$ m. The QCM oscillates
in bulk water;  $f_0$ = $10^7$ Hz.} 

\label{fig7}
\end{figure}
\newpage
\begin{figure}[h]
\centerline{\psfig{figure=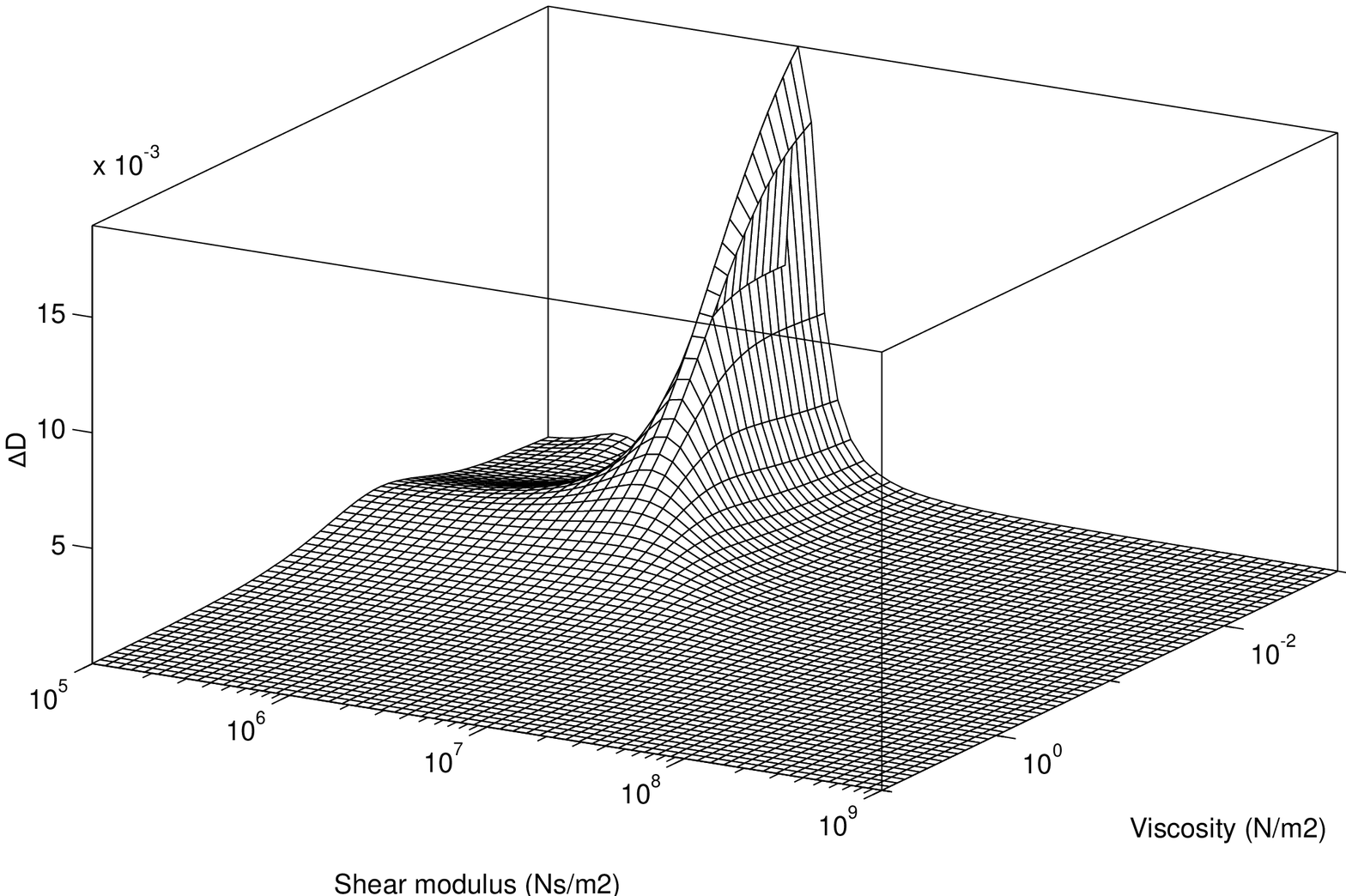,width=15cm}}
\vspace{1.0cm}
\caption{ Numerically calculated dissipation factor change as a function of
the shear elasticity modulus $\mu_1$ and the shear viscosity $\eta_1$ of 
the viscoelastic overlayer of thickness  $h_1 = 10^{-6}$ m.
 The QCM oscillates
 in bulk water; $f_0$ = $10^7$ Hz.} 
          
\label{fig8}
\end{figure}
           

\begin{references}

\bibitem{1}

Thirtle, P. N., Li, Z. X., Thomas, R. K., Rennie, A. R.,
Satija, S. K. and Sung, L. P., Langmuir {\bf 13}, 5451 (1997).

\bibitem{2}
 Richards, R. W., Rochford, B. R., Taylor, M. R., Macromolecules 
{\bf 29}, 980 (1996).

\bibitem{3}
 Fragneto, G., Thomas, R. K., Rennie A. R. and Penfold, J., Science {\bf 267},
657 (1995).

\bibitem{41}
Sauerbrey, G., Z.Phys. {\bf 155}, 206 (1959).

\bibitem{42}
Stockbridge, C. D., in ``Vacuum Microbalance Techniques" (Plenum Press,
{\bf 5}, 147 (1966)).

\bibitem{4}
 Rodahl, M. and Kasemo, B., Sensors \& Actuators {\bf A54}, 448 (1996).

\bibitem{5}
Panizza, P., Roux, D., Vuillaume, V., Lu, C.-Y. D. and Cates, M. E.,
Langmuir {\bf 12}, 248 (1996).

\bibitem{6}
Lvov, Yu., Ariga, K., Ichinose, I. and Kunitake, T., J. Amer. Chem. Soc., 
{\bf 117}, 6117 (1995).

\bibitem{7}
Fredriksson, C., Kihlman, S., Rodahl, M. and Kasemo, B., 
Langmuir (in press) (1997).

\bibitem {8}
Rodahl, M., Hook, F., Fredriksson, C., Keller, C., Krozer, A.,
Brzezinski, P., Voinova, M. V. and Kasemo, B., Faraday Discuss.,  
{\bf 107} (in press) (1997) 

\bibitem {9}
Lucklum, R. and Hauptman, P., ibid.

\bibitem{10}
 Bandley,  H. L., Hillman, A. R., Brown, M. J. and Martin, S. J., ibid.

\bibitem{11}
Voinova, M. V., Jonson, M. and Kasemo, B., J. Phys.: Condens. Matter
  {\bf 9}, 7799 (1997).

\bibitem{12}
Persson, B., ``Sliding Friction" (Springer 1997) (in press).

\bibitem{13}
Kanazawa, K. K. and Gordon J. G., Anal. Chem. {\bf 57} 1770 (1985).

\bibitem{14}
Reed, C. E., Kanazawa, K. K. and Kaufman, J. H., J. Appl. Phys. 
{\bf 68}, 1993 (1990).

\bibitem{15}
Daikhin, L. and Urbakh, M., Langmuir {\bf 12}, 6354 (1996).

\bibitem{16}
Ferry,  J.D., ``Viscoelastic properties of polymers" (3rd ed. New York 1980).

\bibitem{17}
 Philippoff, W. ``Relaxation in Polymer Solutions, Liquids, and Gels",
in: "Physical Acoustics: Principles
 and Methods" {\bf 18} (Edited by W. P. Mason and R. N. Thurston, 
Academic Press New York 1988).

\bibitem{18}
Landau,  L. D., Lifshitz, E. M., ``Fluid Mechanics" (2nd ed., Pergamon,
Oxford, 1987).

\end{references}
\end{document}